\documentclass[11pt]{article}
\usepackage{moriond,epsfig}

\def\Journal#1#2#3#4#5{{#1} {\bf #2}, #3 (#4) {#5}}

 \def\NPB{{\em Nucl. Phys.}
B}  \def\PLB{{\em Phys.  Lett.}   B} 
\def\PRD{{\em Phys. Rev.} D} 

                         \def\be{\begin{equation}}
\def\ee{\end{equation}}                      \def\bea{\begin{eqnarray}}
\def\eea{\end{eqnarray}} 
\begin{document}
\vspace*{4cm}  \title{MEASUREMENT OF  THE  MASSES AND  LIFETIMES OF  B
HADRONS AT THE TEVATRON}

\author{ Pierluigi Catastini \\ (for the CDF and D\O\ collaborations)}

\address{Universita' di Siena and I.N.F.N sez. Pisa, Italy}

\maketitle\abstracts{The latest results for  the $B$ Hadron sector at
the Tevatron  Collider are summarized.  The properties  of $B$ Hadrons
can be  precisely measured at  the Tevatron.  In particularly  we will
focus on the  masses and lifetimes.  The new  Tevatron results for the
$CP$ violation in $B$ Hadrons will be also discussed.}

\section{Introduction}

The  two  experiments, CDF  and  D\O\,  at  the Tevatron  collider  at
Fermilab have  successfully been taking data  with $\sqrt{s}~1.96$ TeV.
New  or  updated  analysis  are   based  on  0.36  -  1  $fb^{-1}$  of
data. Thanks to the possibility of producing all types of $B$ Hadrons,
Tevatron  is exploring  a wide  range  of the  $B$ sector,  performing
precise  measurements of  several properties  like  masses, lifetimes,
production fractions and $CP$ violation.

\section{Measurement of $B$ Hadron relative Fragmentation Fractions}\label{subsec:prod}

In Run  I of the Fermilab  Tevatron, the fraction  of $\bar B^{0}_{s}$
mesons  produced relative  to the  number of  $\bar B^{0}$  mesons was
measured $~2\sigma$ higher at CDF \cite{cdffs} than at LEP experiments
\cite{lep}.To shed light on  the question, recently the CDF experiment
performed the first RUN II measurement of $b$ quark fragmentation into
$\bar  B^{0}$, $B^{-}$, $\bar  B^{0}_{s}$ and  $\Lambda^{0}_{b}$ using
semileptonic decays  reconstructed in 360 $\rm{pb^{-1}}$  of data. The
corresponding fragmentation fractions were measured to be:
\begin{eqnarray}
\frac{f_u}{f_d}&=&1.054\pm   0.018(stat)  ^{+0.025}_{-0.045}(sys)  \pm
0.058({\cal    BR})\\    \frac{f_s}{f_u+f_d}&=&0.160\pm    0.005(stat)
^{+0.011}_{-0.010}(sys)      ^{+0.057}_{-0.034}({\cal      BR})     \\
\frac{f_{\Lambda_b}}{f_u+f_d}&=&0.281\pm0.012(stat)
^{+0.058}_{-0.056}(sys) ^{+0.128}_{-0.086}({\cal BR}).
\end{eqnarray}
Since the measured semileptonic events have an effective $p_T(B)> 7.0$
threshold,  $f_q$  in   this  text  indicates  $f_q\equiv  f_q(p_T(B)>
7.0\mbox{     GeV/c})$.The     relative    fragmentation     fractions
$\frac{f_s}{(f_u+f_d)}$  and  $\frac{f_{\Lambda_b}}{(f_u+f_d)}$ differ
from the world averages by $~1 \sigma$ and $~2 \sigma$ respectively.

\section{Orbitally Excited $B$ Mesons}

The quark  model predicts the  existence of two wide  ($B^{*}_{0}$ and
$B^{*}_{1}$) and two narrow ($B_{1}$ and $B^{*}_{2}$) bound $P$ states
\cite{excitedtheo}.  The wide states  decay via $S$ wave and therefore
have  a  large width  of  a  few  $\rm{MeV/c^{2}}$.  Such  states  are
difficult  to  distinguish from  combinatoric  background. The  narrow
states decay via  $D$ wave and therefore should have  a small width of
around $10 \rm{MeV/c^{2}}$  \cite{excitedtheo}.The same excited states
pattern is predicted for $B^{0}$, $B^{+}$ and $B_{s}$ mesons.

\subsection{B excited states}\label{bstar}

\begin{figure}
\psfig{file=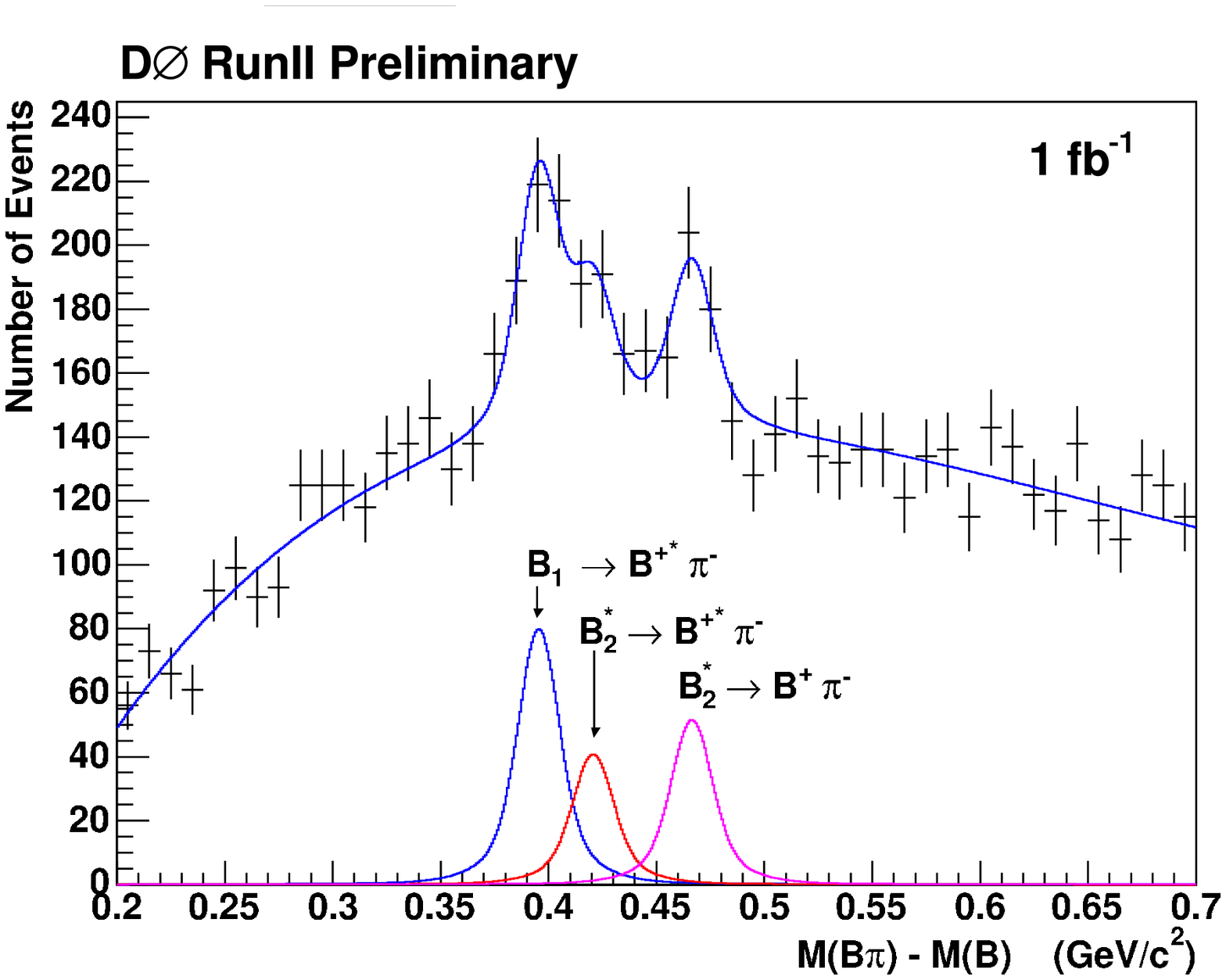,width=2.9in,height=1.8in}
\psfig{file=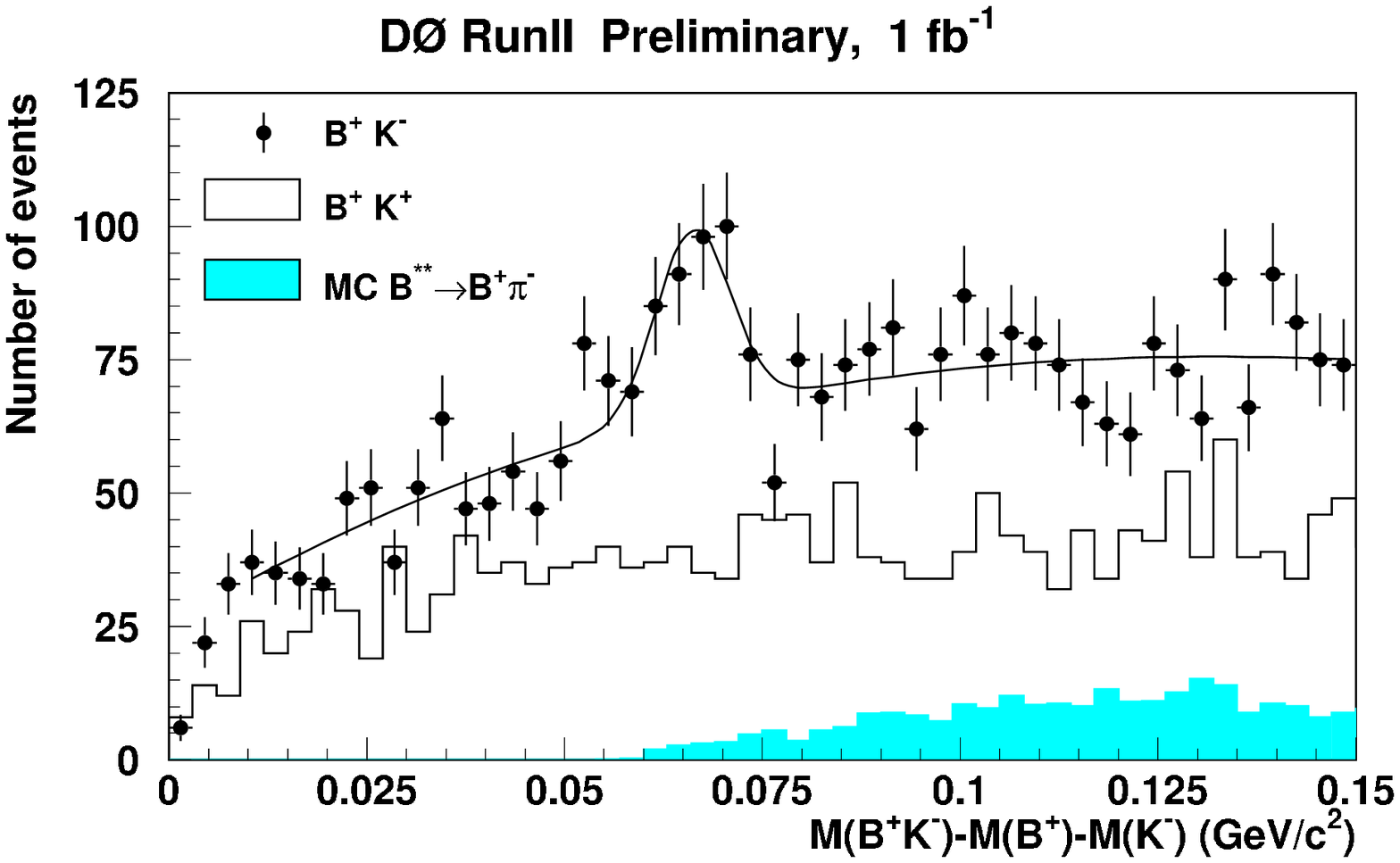,width=2.9in,height=1.8in}
\caption{Left: $M(B\pi)  - M(B)$ distribution. Right: $M(BK)  - M(B) -
M(K)$ distribution.}
\label{fig:bstar}
\end{figure}

Recently the D\O\ experiment  studied the narrow $L=1$ states decaying
to $B^{+(*)} \pi$  with exclusively reconstructed $B\rightarrow J/\psi
K$ in  1 $ \rm{fb^{-1}}$  of data.  A  clear excess of events,  with a
statistical significance of more than $~7 \sigma$, can  be observed in
the $M(B\pi) - M(B)$  distribution of Fig.\ref{fig:bstar} (left). This
excess can be interpreted in terms of three contributions:
\begin{eqnarray}
\label{eq:bstardecay}
\nonumber{B^{0}_{1}   \rightarrow   B^{*+}\pi^{-};   B^{*+}\rightarrow
B^{+}\gamma}\\    \nonumber{B^{*0}_{2}    \rightarrow   B^{*+}\pi^{-};
B^{*+}\rightarrow  B^{+}\gamma}  \\  \nonumber{B^{*0}_{2}  \rightarrow
B^{+} \pi^{-}; \quad \quad \quad \quad \quad \quad}
\end{eqnarray}
where the  $\gamma$ from $B^{*+}$ is not  reconstructed (that explains
the separation  between the two $B^{*0}_{2}$  decays). The $B^{0}_{1}$
and  $B^{*0}_{2}$  masses  and  width  (same  width  is  assumed)  are
measured:
\begin{eqnarray}
\label{eq:bstarmass}
\nonumber{M(B_{1}) =  5720.8 \pm 2.5  (stat.)  \pm 5.3  ~(syst.) \quad
MeV/c^{2}}\\ \nonumber{M(B^{*}_{2}) - M(B_{1}) = 25.2 \pm 3.0 ~(stat.)
\pm 1.1 ~(syst.) \quad MeV/c^{2}} \\ \nonumber{\Gamma_{1}=\Gamma_{2} =
6.6 \pm 5.3 ~(stat.) \pm 4.2 ~(syst.) \quad MeV/c^{2}}
\end{eqnarray}
while the branching ratio  of $B^{*0}_{2}$ to $B^{*}$, the composition
of the  $B_{j}$ sample and  the $B_{j}$ production rate  normalized to
the $B^{+}$ one are measured to be:
\begin{eqnarray}
\label{eq:bstarratio}
\nonumber{\frac{Br(B^{*}_{2}\rightarrow
B^{*}\pi)}{Br(B^{*}_{2}\rightarrow  B^{(*)}\pi)}  =  0.513  \pm  0.092
~(stat.)  \pm  0.115 ~(syst.)}  \\ \nonumber{\frac{Br(B_{1}\rightarrow
B^{*+}\pi)}{Br(B_{j}\rightarrow   B^{(*)}\pi)}  =   0.545   \pm  0.064
~(stat.)   \pm  0.071  ~(syst.)}   \\  \nonumber{\frac{Br(b\rightarrow
B^{0}_{j}\rightarrow B\pi)}{Br(b\rightarrow B^{+})}  = 0.165 \pm 0.024
~(stat.) \pm 0.028 ~(syst.)}
\end{eqnarray}

\subsection{$B^{0}_{s}$ excited states}

Using  a  sample of  1  $\rm{fb^{-1}}$  of  data the  D\O\  experiment
observed the process $B^{*0}_{2s}\rightarrow  B^{+}K$ An excess with a
statistical  significance  exceeding  $5\sigma$  can  be  observed  in
Fig.\ref{bstar} (right). The $B^{*0}_{2s}$ mass is measured:
\begin{eqnarray}
\label{eq:bstar}
\nonumber{M(B^{*0}_{s2}) =  5839.1 \pm  1.4 ~(stat.) \pm  1.5 ~(syst.)
\quad MeV/c^{2}}
\end{eqnarray}
Taking into account  that the theory predicts the  same mass splitting
in  the  $\bar{b}d(u)$  and  $\bar{b}s$ systems  and  considering  the
results shown in section \ref{bstar}, the decay $B^{0}_{s1}\rightarrow
B^{*+}K$   is  forbidden   (it   is  under   decay  threshold)   while
$B^{*0}_{s2}\rightarrow  B^{+}K$ is  strongly suppressed  due  to phase
space.

\section{$B_{c}$ meson properties}

\begin{figure}[!h]
\psfig{file=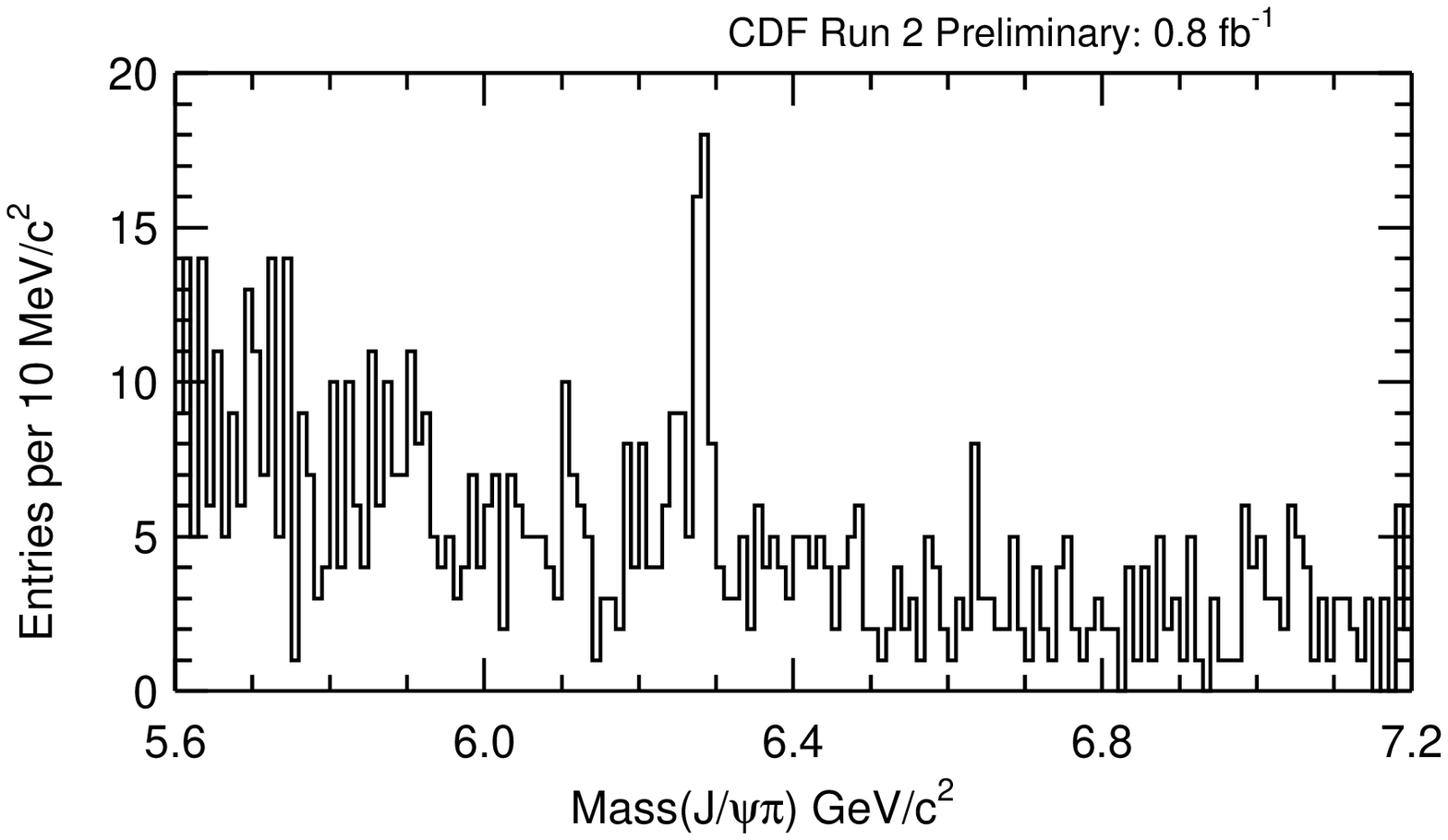                        ,width=2.9in,height=1.8in}
\psfig{file=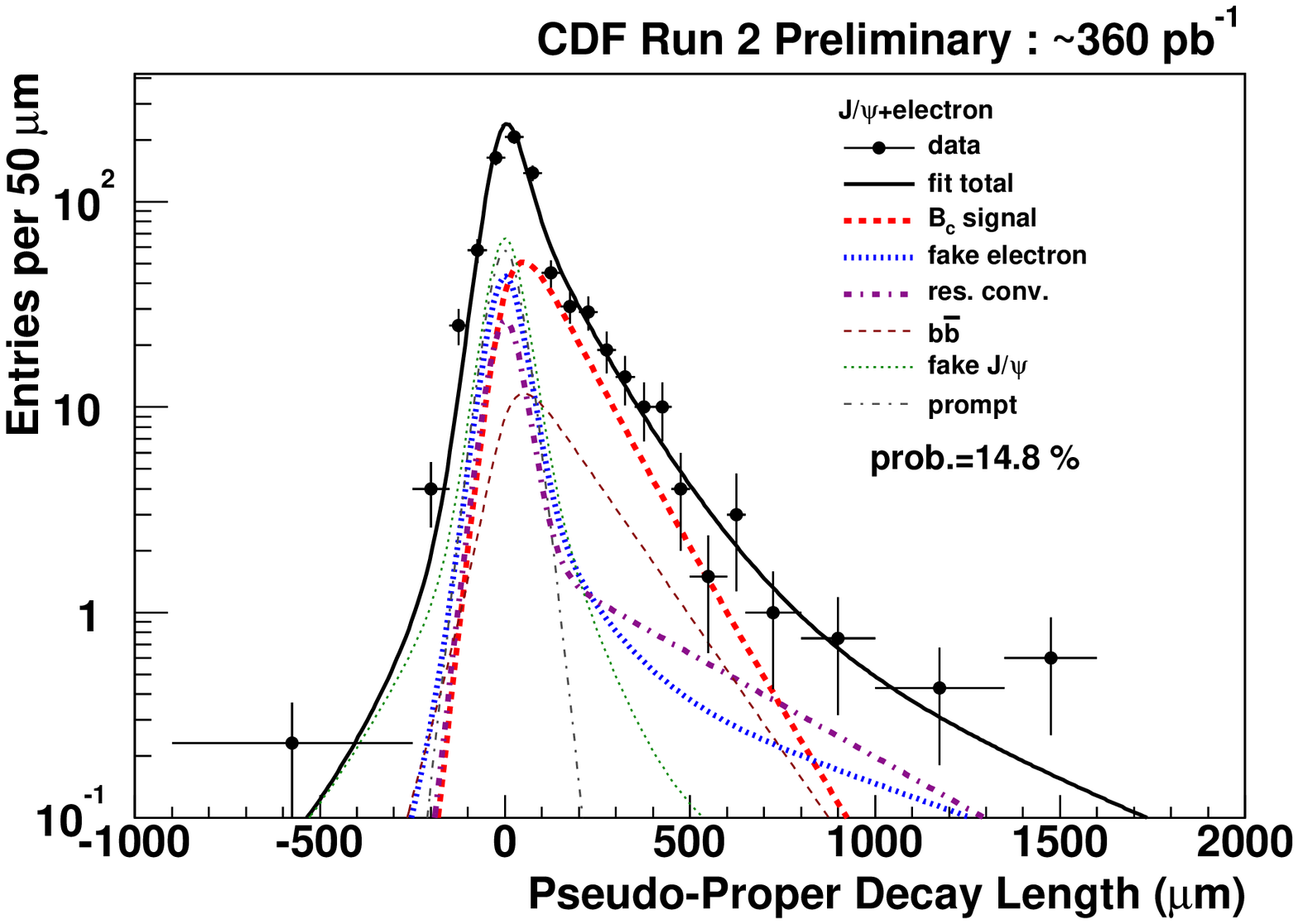,width=2.9in,height=1.8in}
\caption{Left: invariant mass  distribution of $B_{c}\rightarrow J\psi
\pi$  candidates. Right:  pseudo-proper decay  length  distribution of
$B_{c}\rightarrow   J/\psi  e   \nu$  events   overlaid  to   the  fit
projection.}
\label{fig:bc}
\end{figure}

Using  a sample  of  $0.8 \rm{fb^{-1}}$  the  CDF experiment  recently
updated the $B_{c}\rightarrow J\psi \pi$ exclusive decay analysis.  In
Fig.\ref{fig:bc}  an excess,  with a  statistical  significance greater
than $6\sigma$,  can be  observed. The new  measurement of  the $B_{c}$
mass is:
\begin{eqnarray}
\label{eq:bcmass}
\nonumber{M(B_{c}) =  6275.2 \pm 4.3  ~(stat.) \pm 2.3  ~(syst.) \quad
MeV/c^{2}}
\end{eqnarray}
Using the decay $B_{c}\rightarrow  J/\psi e \nu$ in $360 \rm{pb^{-1}}$
of  data (in  Fig.\ref{fig:bc}  right the  pseudo-proper decay  length
distribution), CDF measured also the $B_{c}$ lifetime to be:
\begin{eqnarray}
\label{eq:bclife}
\nonumber{\tau(B_{c}) =  0.474 ^{+0.073}_{-0.066} ~(stat.)   \pm 0.033
~(syst.)\quad ps}
\end{eqnarray}

\section{CP violation parameter in $B^{0}$ mixing and decay}

The  CP-violation  parameter (\ref{eq:epsib}) is sensitive to  several
extensions of the Standard Model  \cite{cppar} and it is considered as
one of the  first possible indications of physics  beyond the Standard
Model. The D\O\ experiment, using $~970 pb^{-1}$ of data, measured the
dimuon charge asymmetry.  As a  sensitive cross check, the mean mixing
probability  $<\chi>$ of hadrons with a $b$ quark was also measured:
\begin{eqnarray}
\label{eq:chi}
\nonumber{<\chi> = 0.136 \pm 0.001 ~(stat.) \pm 0.024 ~(syst.)}
\end{eqnarray}
Assuming that the dimuon asymmetry $A_{B^{0}}$, i.e. the dimuon charge
asymmetry from  direct-direct decays of $B^{0}  \bar{B^{0}}$ pairs, is
due to asymmetric $B^{0}\leftrightarrow \bar{B^{0}}$ mixing decay, the
CP-violation parameter of $B^{0}$ mixing and decay was extracted. The
corresponding measurements were found to be:
\begin{eqnarray}
\label{eq:epsib}
\nonumber{A_{B^{0}}   =  0.0044  \pm   0.0040  ~(stat.)    \pm  0.0028
~(syst.)}\\    \frac{Re(\epsilon_{B^0})}{1+|\epsilon_{B^{0}}|^{2}}   =
\frac{A_{B^{0}}}{4} = -0.0011 \pm 0.0010 ~(stat.) \pm 0.0007 ~(syst.)
\end{eqnarray}

\section{CP violation and $\Delta\Gamma_{\mathsf{CP}}/ \Gamma_{\mathsf{CP}}$ in B charmless two body decays}

\begin{figure}
\psfig{file=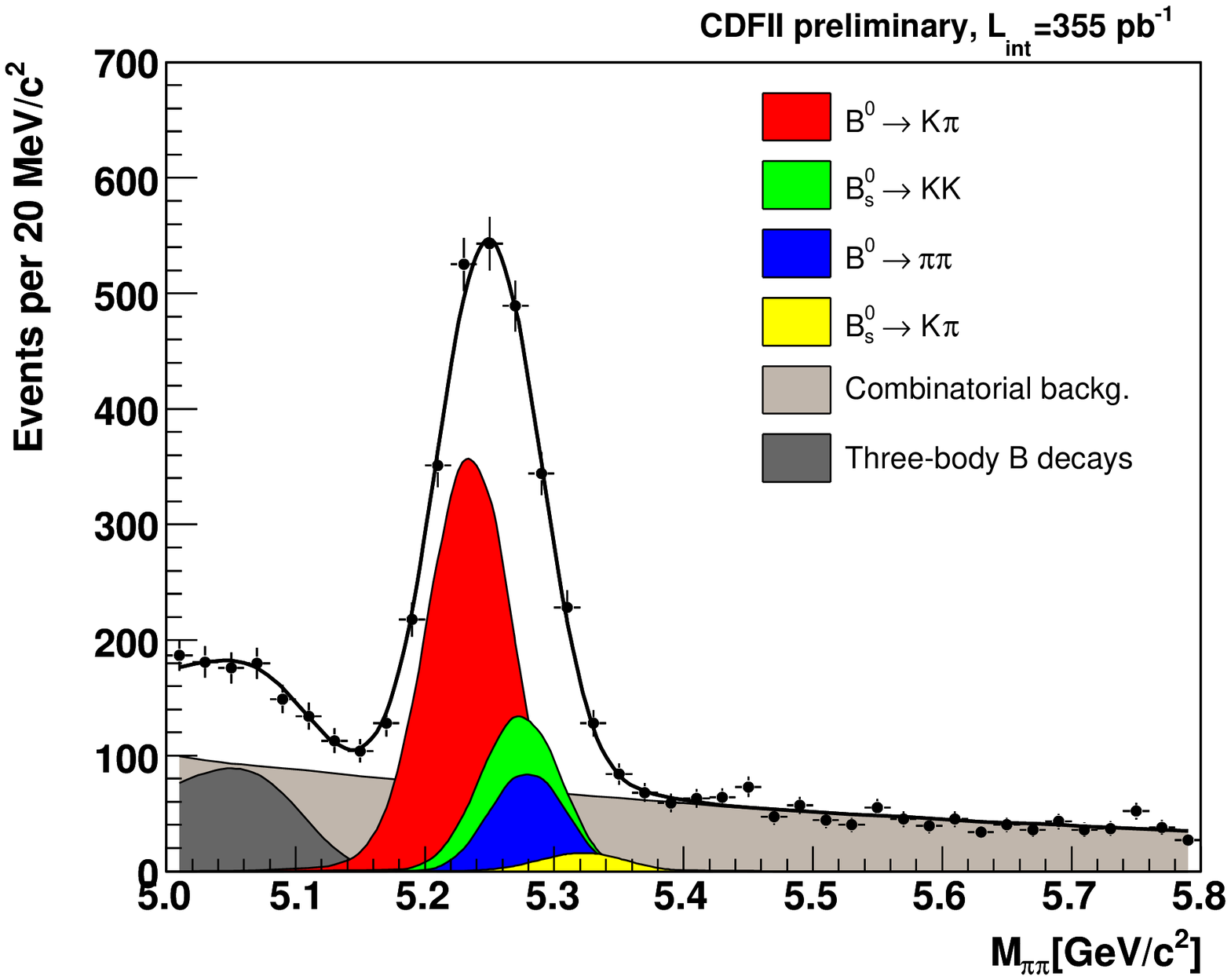,width=2.9in,height=1.8in}
\psfig{file=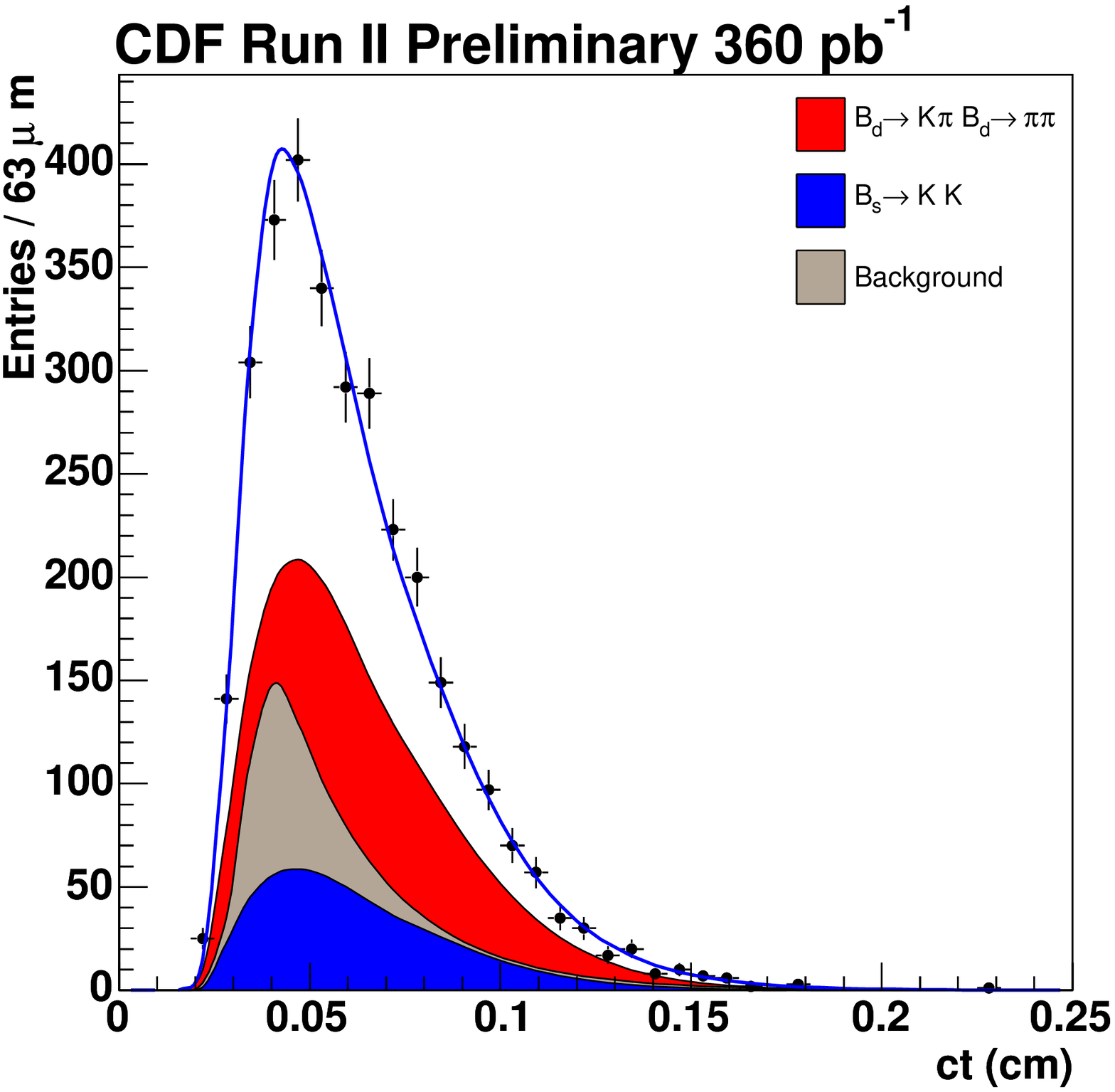 ,width=2.9in,height=1.8in}
\caption{Right:   invariant   mass   distribution   of   $B\rightarrow
h^{+}h^{-}$ candidates  overlaid to  the fit projection.  Right: proper
decay length distribution overlaid to the fit projection.}
\label{fig:bhh}
\end{figure}

The CDF II detector can collect and reconstruct significant samples of
$B^{0}$ and  $B^{0}_{s}$ charmless two  body decays, Fig.\ref{fig:bhh}
(left).   Recently  CDF  updated  his  measurement of  the  direct  CP
asymmetry  in  $B^{0}\rightarrow  K^{+}\pi^{-}$  events  using  $~$355
$\rm{pb^{-1}}$ of data. The new measurement was found to be:
\begin{eqnarray*}
\label{eq:acp}
A_{\mathsf{CP}}   =   \frac{N(\overline{B}^0\rightarrow  K^-\pi^+)   -
N(B^0\rightarrow  K^+\pi^-)}{N(\overline{B}^0\rightarrow  K^-\pi^+)  +
N(B^0\rightarrow   K^+\pi^-)}   =   -0.058  \pm   0.039~(stat.)    \pm
0.007~(syst.)
\end{eqnarray*} 
Using  the  same data  sample,  the $B^{0}$  two  body  decay and  the
$B^{0}_{s}\rightarrow   K^{+}K^{-}$  lifetimes   were   measured.  The
corresponding values were found to be:
\begin{eqnarray}
\label{eq:bskk}
\nonumber{\tau(B^{0}) = 1.51 \pm  0.08 ~(stat.) \pm 0.02 ~(syst.)\quad
ps}  \\ \nonumber{\tau(B_{s}\rightarrow  K^{+}K^{-}) =  1.53  \pm 0.18
~(stat.) \pm 0.02 ~(syst.)\quad ps}
\end{eqnarray}
lifetime fit projections are shown in Fig.\ref{fig:bhh} (right).

Combining  the   CDF  II  measurement   of  the  $B^{0}_{s}\rightarrow
K^{+}K^{-}$ lifetime with the latest HFAG average $B^{0}_{s}$ lifetime
in flavor  specific decays  $\tau(B_{s} FS) =  1.454 \pm  0.040$ $ps$,
$\Delta\Gamma_{\mathsf{CP}}/  \Gamma_{\mathsf{CP}}$  was evaluated  to
be: \cite{hfag}:
\begin{eqnarray}
\label{eq:dgamma}
\nonumber{\frac{\Delta
\Gamma_{\mathsf{CP}}}{\Gamma_{\mathsf{CP}}}(B_{s}\rightarrow
K^{+}K^{-}) = -0.08 \pm 0.23 ~(stat.) \pm 0.03 ~(syst.)}
\end{eqnarray}

%
%
%

\end{document}